\documentclass[12pt, draftclsnofoot, onecolumn]{IEEEtran}
\normalsize

\usepackage{amsmath,amsfonts,amssymb,amsbsy,verbatim}
\usepackage{pgfplots}
\pgfplotsset{compat=1.16}
\usepackage{times}
\usepackage{booktabs}
\usepackage{dsfont}
\usepackage{cite}
\usepackage{bm}
\usepackage{epstopdf}
\usepackage{algorithm,algorithmic, color}
\usepackage{todonotes}
\presetkeys{todonotes}{inline}{}
\usepackage{amsthm}
\usepackage{bm}
\usepackage{graphicx}
\usepackage{sparklines}
\usepackage{tikz}
\usepackage{balance}
\usetikzlibrary{positioning}
\usetikzlibrary{calc}
\usetikzlibrary{shapes.geometric}
\usetikzlibrary{decorations.pathreplacing} 
\newcommand{\cloud}[3]{
\begin{scope}[shift={#1},scale=#3]
\draw[fill=white] (-1.6,-0.7) .. controls (-2.3,-1.1)
and (-2.7,0.3) .. (-1.7,0.3)coordinate(asy1) .. controls (-1.6,0.7)
and (-1.2,0.9) .. (-0.8,0.7) .. controls (-0.5,1.5)
and (0.6,1.3) .. (0.7,0.5) .. controls (1.5,0.4)
and (1.2,-1) .. (0.4,-0.6)coordinate(asy2) .. controls (0.2,-1)
and (-0.2,-1) .. (-0.5,-0.7) .. controls (-0.9,-1)
and (-1.3,-1) .. cycle;
\node at ($(asy1)!0.5!(asy2)$) {#2};
\end{scope}
}
\tikzset{
    >=latex,
    trapezium stretches=true
}

\makeatletter
\let\MYcaption\@makecaption
\makeatother
\usepackage[font=footnotesize,labelformat=simple]{subcaption}
\makeatletter
\let\@makecaption\MYcaption
\makeatother

\usepackage[T1]{fontenc}
\usepackage{stmaryrd}

\newcommand{\bi}{\begin{itemize}}
\newcommand{\ei}{\end{itemize}}
\newcommand{\ben}{\begin{enumerate}}
\newcommand{\een}{\end{enumerate}}
\newcommand{\bc}{\begin{cases}}
\newcommand{\ec}{\end{cases}}
\newcommand{\bd}{\begin{description}}
\newcommand{\ed}{\end{description}}

\newcommand{\be}{\begin{equation}}
\newcommand{\ee}{\end{equation}}
\newcommand{\bea}{\begin{eqnarray}}
\newcommand{\eea}{\end{eqnarray}}

\pgfplotsset{compat=1.13} 
\definecolor{col0}{rgb}{0, 0, 0}
\definecolor{col1}{rgb}{0.0000,0.4470,0.7410}%
\definecolor{col2}{rgb}{0.8500,0.3250,0.0980}%
\definecolor{col3}{rgb}{0.9290,0.6940,0.1250}%
\definecolor{col4}{rgb}{0.4940,0.1840,0.5560}%
\definecolor{col5}{rgb}{0.4660,0.6740,0.1880}%
\definecolor{col6}{rgb}{0.3010,0.7450,0.9330}%

\pgfplotsset{compat=1.16,
    /pgfplots/ybar legend/.style={
    /pgfplots/legend image code/.code={%
       \draw[##1,/tikz/.cd,yshift=-0.25em]
        (0cm,0cm) rectangle (3pt,0.8em);},
   },
}

 \pgfdeclareplotmark{xo}{%
  \pgfusepathqstroke
  \pgfnode{rectangle}{center}{$\star$}{}{}
}

\IEEEoverridecommandlockouts

\begin{document}
\title{Triplet-Based Wireless Channel Charting: Architecture and Experiments}
\author{Paul~Ferrand, Alexis~Decurninge, Luis~G.~Ordoñez, Maxime~Guillaud\\
Mathematical and Algorithmic Sciences Laboratory, Paris Research Center, Huawei Technologies France
\thanks{This article builds upon preliminary work on triplet-based channel charting presented in the conference article~\cite{Charting_Globecom}.}
}
\maketitle

\begin{abstract}
Channel charting is a data-driven baseband processing technique consisting in applying self-supervised machine learning techniques to channel state information (CSI), with the objective of reducing the dimension of the data and extracting the fundamental parameters governing its distribution.
We introduce a novel channel charting approach based on triplets of samples.
The proposed algorithm learns a meaningful similarity metric between CSI samples on the basis of proximity in their respective acquisition times, and simultaneously performs dimensionality reduction.
We present an extensive experimental validation of the proposed approach on data obtained from a commercial Massive MIMO system; in particular, we evaluate to which extent the obtained channel chart is similar to the user location information, although it is not supervised by any geographical data.
Finally, we propose and evaluate variations in the channel charting process, including the partially supervised case where some labels are available for part of the dataset.
\end{abstract}


\section{Introduction}

Recent evolutions of wireless network standards involve an increasingly high numbers of antennas on the access points or base stations (BS), and have expanded the bandwidth used for communication~\cite{Larsson2014}.
This evolution comes at the expense of an increase in the number of parameters (known collectively as channel state information, or CSI) necessary to track the propagation channel in real-time by the BS for all users in the system, and poses a number of technical challenges.
In particular, it makes it difficult for a given BS to predict the evolution of its users' channels over time, and therefore allocate resources in an optimal way.

\subsection{Channel Charting}
\label{sec:dimensionality_reduction}
Radio propagation models indicate that, for a given physical location (e.g. a particular BS and the surrounding scattering environment), the state of the radio channel should essentially be governed by a manageable number of parameters---namely, the geometry, positions and orientations of the antenna arrays, and of the main scatterers.
If one considers CSI as a random variable (typically of high dimension), and discounts temporal correlation, it is intuitively clear that the distribution of the CSI process is non-uniform; for instance, some realizations will occur extremely unfrequently, because the corresponding channel impulse responses cannot physically be realized in the considered scattering environment.
Departing from the traditional geometry-based channel models, channel charting \cite{Studer_etal_charting} is a data-driven attempt to identify the \emph{information geometry} of CSI, i.e. the shape and properties of the CSI distribution, without relying on classical propagation models.

Channel charting consists in the application of self-supervised dimensionality reduction methods from the field of machine learning to CSI data; see e.g. \cite{McInnes2018} or\cite{Lee2007} for an overview of dimensionality reduction methods.
Based on a large CSI dataset acquired in a given propagation environment, it maps each CSI sample (belonging to the high-dimensional \emph{ambient} space) to a point in a \emph{latent} space of lower dimension, while attempting to preserve the relative sample distances between the ambient and latent spaces.
The image of the dataset in the latent space is referred to as a \textit{chart}.
Due to its self-supervised nature (the coordinates in the latent space are not deemed to represent any meaningful quantity), channel charting is applicable without any knowledge of the user location nor of a faithful geometric model of the environment (e.g., involving the location, orientation and geometry of the involved antennas or antenna arrays).
As such, the chart is an attempt to identify a set of essential parameters capturing the information contained in the CSI.
Desirable properties of a chart are: the mapping should remain consistent in time and across users, and the CSI distribution should map to a simple (e.g. compact) distribution in the latent space.
In~\cite{Studer_etal_charting} and subsequent works~\cite{Deng2018,lei2019siamese,huang2019charting_representation_constrained}, several channel charting approaches have been introduced, centered around Sammon's mapping, autoencoders, and the so-called Siamese networks.

A channel chart can be used to enhance numerous network functionalities, such as predictive radio resource management and rate adaptation, handover between cells, beam association and user tracking and pairing or grouping in device-to-device scenarios.
In particular, the UE position in space is a predominant parameter, as demonstrated by numerous works on CSI-based localization (see \cite{WEN_etal_localization_survey_DSP2019} and references therein), and using location to predict the channel quality of users is at the heart of the idea of radio maps \cite{Taranto_etal_location_aware_2014, Bui_etal_anticipatory_2017}.
In a number of location-based applications, it is possible to use the channel chart as a replacement for a radio map (i.e. to replace the user's absolute position by a pseudo-position on a chart).

\subsection{Metric Learning}

Textbook dimensionality reduction is based on the principle of neighborhood preservation between the ambient and the latent space, i.e. distances between two samples in the ambient space should be preserved in the latent space whenever it is small.
Typically, this is formulated using the Euclidean distance on both spaces.
However, the Euclidean distance is notoriously unreliable in high dimension, and especially when applied to CSI data due to the high dynamic range of the samples, as well as the acquisition procedure which introduces spurious perturbations typically not captured in standard channel models---including imperfect synchronization, timing advance, automatic gain control and transmit power control, and manufacturing variability between nominally identical devices.
So far, this issue was addressed in part through feature engineering, whereby a feature space is used as the ambient space of the dimensionality reduction problem, and the CSI-based features are designed using expert knowledge.
For instance in~\cite{Studer_etal_charting}, the features are chosen as the second-order moments of the CSI coefficients, while in~\cite{Deng2018} the features are obtained using a multi-path component decomposition of the channel impulse response in the spatial and temporal domain.
In both cases, the geographical distance directly appears in the feature model.

Conversely, in the present work, the preprocessing step mostly aims at cleaning the CSI data from the sampling impairements, and we use metric learning to infer a meaningful distance between the preprocessed CSI samples.
Metric learning originates from classification problems, where the objective is to learn a distance that has a good discrimination capability between same-class and inter-class samples. 
Classical works on metric learning include~\cite{Weinberger2009,NIPS2009_3658_BoostMetric}, and are based on sample \emph{triplets}, whereby 2 elements of the triplets belong to the same class, while the last element is an outsider; in these works, the distance is constrained to be a Mahalanobis distance.
More recently, a similar approach has been proposed for face classification under the name ``FaceNet''~\cite{Schroff2015}, using a deep neural network as the basis of a parametric representation of the distance.

\subsection{Contributions}

In this article, we consider deep neural network (DNN)-based dimensionality reduction, i.e. we seek to develop algorithmic approaches capable of identifying a set of possibly non-linear functions modeled by a neural network, that allow to parameterize the empirical CSI distribution observed by a given BS using a small number of parameters.
Our contribution is as follows:
\begin{itemize}
\item We review the triplet-based channel charting method we introduced in~\cite{Charting_Globecom}, which performs joint dimensionality reduction and CSI distance learning. 
The proposed approach builds on the idea that a meaningful distance can be learned solely from sample proximity in time.
Since the sampling epoch (or time-stamp) information is easily available in the context of CSI acquisition, this constitutes a weak form of supervision.
\item We discuss several classical quality metrics used for dimensionality reduction methods, and their applicability in the context of channel charting.
\item We validate experimentally the proposed charting method on a massive MIMO CSI dataset acquired on a commercial 4G BS, and benchmark the proposed approach against classical dimensionality reduction methods and previous channel charting approaches in the literature~\cite{Studer_etal_charting,McInnes2018,Deng2018,lei2019siamese,huang2019charting_representation_constrained}.
Furthermore, in order to evaluate to which extent channel charts can provide relative location information, we compute similarity metrics between chart and user position for this dataset.
\item We also report on various extensions of the approach including curriculum learning, variation in the DNN structure, and semi-supervised training of the channel chart.
\end{itemize}

The remainder of this paper is organized as follows.
We describe the dimensionality reduction problem in a general form in Section~\ref{sec:problem_description}, presenting the generative model we assume in order to extract a meaningful chart from the dataset.
In Section~\ref{sec:sota_charting}, we cover state-of-the-art channel charting approaches, and discuss their suitability for the problem at hand.
Our triplet-based approach to channel charting is detailed in Section~\ref{sec:triplets}.
In Section~\ref{sec:experimental_results} we describe the experimental data used for validation, and the data processing pipeline.
Finally, we evaluate some extensions of the triplet-based channel charting in Section~\ref{sec:extensions} and their relative merits in improving the performance of our original approach.

\section{Dimensionality Reduction}
\label{sec:problem_description}

To ground our subsequent discussion onto a proper mathematical framework, let $X$ be a random variable of unknown distribution with values in $\mathcal X \subset \mathbb R^D$.
Assume that $N$ samples from $X$ are available and indexed by $i \in \mathcal S$; $\mathcal S$ has cardinality $|\mathcal S| = N$.
Given another space $\mathcal Y \subset \mathbb R^d$ with $d < D$, classical dimensionality reduction consists in assigning coordinates $\bm y_i \in \mathcal Y$ to the $i$-th sample of $X$ (denoted by $\bm x_i$), such that
\begin{equation}
    \label{eq:distance_equality}
    \|\bm y_i - \bm y_j\| \approx d(\bm x_i, \bm x_j), \qquad \text{\ for \ } (i,j) \in \mathcal T \subset \mathcal S^2
\end{equation}
where $\|\cdot\|$ denotes the Euclidean norm on the latent space $\mathbb R^d$, and $d(\cdot, \cdot)$ denotes a distance in the ambient space $\mathbb R^D$.
Throughout this article, $\mathcal T$ denotes a set of sample indices; depending on the context it will be pairs or triplets of indices.
In the context of wireless communications, the sample $\bm x_i$ is usually the result of preprocessing the CSI $\bm h_i$ in order to clean the data and facilitate the construction of meaningful features from the latter.
It is not generally possible to achieve equality in \eqref{eq:distance_equality} unless the dimension of the latent space is very large, and dimensionality reduction algorithms merely aim to minimize the violation of the equality constraint.
The above problem formulation generalizes multiple existing approaches:
\begin{itemize}
  \item Taking $\mathcal T = \mathcal S^2$ together with the Euclidean distance on $\mathcal X$ (i.e. $d(\bm x, \bm x') = \|\bm x' - \bm x\|$) yields the classical multi-dimensional scaling (MDS) approach~\cite{Kruskal1964, Lee2007}.
  \item Sammon's mapping~\cite{Sammon1969, Lee2007} casts~\eqref{eq:distance_equality} as an optimization problem, still considering $\mathcal T = \mathcal S^2$ and taking $d(\cdot, \cdot)$ as the Euclidean distance. The importance of the constraint in~\eqref{eq:distance_equality} is weighted by a factor monotonically decreasing in $\|\bm x_i - \bm x_j\|$.
  \item Taking $\mathcal T$ as the set of indices pairs $(i,j)$ such that $\bm x_i$ and $\bm x_j$ are neighbors in $\mathcal X$ (e.g., with $\| \bm x_i - \bm x_j\|$ below a threshold) is the starting point of the Isomap algorithm \cite{Tenenbaum2000, Lee2007}. The distance $d(\cdot, \cdot)$ in the Isomap algorithm is then computed from the graph of neighbor samples as the shortest path in the graph.
\end{itemize}

Let us assume for now that $d(\cdot, \cdot)$ is known; this matter is discussed further in Section~\ref{sec:triplets}.
Of particular interest is the ideal case where there exists a generative model such that $X$ arises from a $d$-dimensional random variable $W$ with values in $\mathcal W \subset \mathbb R^d$ via a mapping $g : \mathcal W \rightarrow \mathcal X$, i.e.,
\begin{equation}
    \label{eq:basic_model}
    \bm x_i = g(\bm w_i),  \qquad\text{\ for \ } i \in \mathcal S 
\end{equation}
where $\bm w_i$ is sampled from $W$.
In that case, $g(\mathcal W)$ is a manifold embedded in $\mathbb R^D$, and $X$ is a manifold-valued random variable.
Assuming that $g$ is an homeomorphism, choosing $\bm y_i \propto g^{-1}(\bm x_i) = \bm w_i$ fulfills \eqref{eq:distance_equality} with equality when $\mathcal T$ is chosen to contain infinitesimally small neighborhoods of samples.
When dimensionality reduction is applied to CSI samples, $W$ can be related to the parameters of the Maxwell equations that govern propagation, while $X$ is the channel state, and $\mathcal Y$ is the channel chart.  

Generally, however, the generative model (the function $g$ and the $\bm w_i$) is unknown, and only the observations $\bm x_i$ are available.
Therefore, one seeks to infer a mapping $f : \mathcal X \rightarrow \mathcal Y$ such that the $\bm y_i$ are given by $\bm y_i = f(\bm x_i)$ and the distance preservation property \eqref{eq:distance_equality} is enforced (at least approximately), i.e., for all $i,j$, it holds $\|f(\bm x_i) - f(\bm x_j)\| \approx d(\bm x_i, \bm x_j)$.
Therefore, $f$ and $\mathcal Y$ act as surrogates for $g^{-1}$ and $\mathcal W$ when the generative model is unknown.
This can be summarized as follows:
\begin{equation}
\begin{array}{ccc}
    \underbrace{
    \begin{array}{cc}
        \mathcal W \subset \mathbb R^d & \rightarrow \\
        \bm w_i & \stackrel{g}{\mapsto} 
    \end{array}
    }_{\text{Generative model}} & \underbrace{
    \begin{array}{c}
         \mathcal X \subset \mathbb R^D  \\
         \bm x_i 
    \end{array}
    }_{\text{Observations}} & \underbrace{
    \begin{array}{cc}
        \rightarrow  & \mathcal Y \subset \mathbb R^d  \\
        \stackrel{f}{\mapsto} & \bm y_i
    \end{array}
    }_{\text{Dimensionality Reduction}}
\end{array}
\end{equation}
Note that the problem of finding $f$ can only be solved up to an isometric transformation of $\mathcal Y$, since it is entirely defined by the distance constraints \eqref{eq:distance_equality}.
Furthermore, since $d$ is usually not known in practice, a heuristic choice of the dimension of $\mathcal Y$ is required.

\subsection{DNN-Based Dimensionality Reduction}
The focus of the rest of this article is on DNN-based approaches to dimensionality reduction.
In this context, $f$ is constrained to be a parametric function defined by a DNN, which is trained to optimize an objective function related to the dimensionality reduction objective.
Letting $\bm \theta$ denote the parameters (weights) of the DNN and $f_{\bm \theta}$ its transfer function, the embedding of the dataset in $\mathcal Y$ is obtained as the image of  $\mathcal S$ by $f_{\bm \theta}$, namely
\begin{equation}
  \label{DNN_mapping}
   \bm y_i = f_{\bm \theta}(\bm x_i) \quad \text{for } i \in \mathcal S.
\end{equation}
The optimal value of $\bm \theta$ is determined from the training dataset according to
\begin{equation}\label{eq_DR_thetastar}
  \bm \theta^* =  \arg\min_{\bm \theta} \frac{1}{|\mathcal T|} \sum_{(i, j) \in \mathcal T}  \Big(\| f_{\bm \theta}(\bm x_i) - f_{\bm \theta}(\bm x_j)\| - d(\bm x_i, \bm x_j) \Big)^2,
\end{equation}
and $f$ is chosen as $f=f_{\bm \theta^*}$. 
Clearly, the above formulation promotes \eqref{eq:distance_equality}.
The DNN-based parametric approach has a significant advantage over other dimensionality reduction approaches: it handles gracefully the so-called \emph{out-of-sample} problem, i.e., once $\bm \theta^*$ has been determined, the chart coordinates for a new sample $\bm x$ not present in $\mathcal S$ can be obtained easily by evaluating $f_{\bm \theta^*}(\bm x)$.
Other methods based on DNNs, mainly autoencoders or Siamese networks also enjoy this desirable property.
This is in contrast with some of the state-of-the-art methods such as ISOMAP, Sammon's mapping or UMAP, which consist in jointly computing the coordinates $\bm y_i$ for each $i$ in the dataset, and therefore would require to solve the whole problem again or to apply barycentric methods~\cite{Bengio2004}, usually at much higher computational cost, in order to handle new samples.

\section{Channel Charting State-of-the-Art}
\label{sec:sota_charting}
Among the channel charting approaches considered in the literature, autoencoders \cite{huang2019charting_representation_constrained} and Siamese networks \cite{lei2019siamese} have received particular attention.
Let us first describe these strategies, and then show how the proposed architecture departs from them.
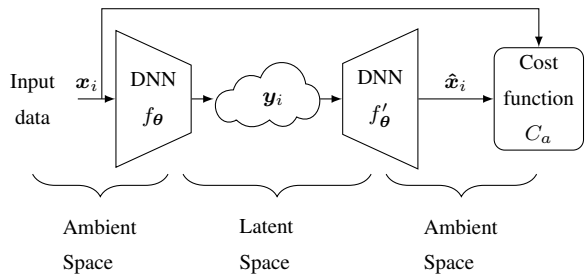
\begin{figure}[ht!]
    \centering
    \scriptsize
    \begin{tikzpicture}[
        trapz/.style = { trapezium, draw, minimum height=1cm, align=center},
        lab/.style = { text width=1cm, text centered },
        minus/.style = { circle, draw },
        box/.style = { draw, rounded corners, text centered }
    ] 
        \node[lab] (training) at (-5, 0) {Input\\data};
        \node[trapz, shape border rotate=270, right=0.5 of training] (reduce) {DNN\\$f_{\bm \theta}$};
        \cloud{(-1.65, 0)}{}{0.4};
        \node at (-1.8, 0) {$\bm y_i$};
        \node[trapz, shape border rotate=90, right=2 of reduce] (expand) {DNN\\$f'_{\bm \theta}$};
        \node[box, lab, right=1 of expand] (cost) {Cost\\function $C_a$};
        \draw[->] (training) -- (reduce) node [near start, above] {$\bm x_i$};
        \draw[->] (reduce) -- ++(0.8, 0);
        \draw[->] (expand) ++(-0.8, 0) -- (expand);
        \draw[->] (expand) -- (cost) node [midway, above] {$\bm{\hat x}_i$};
        \draw[->] (training) ++(0.92, 0) -- ++(0, 1.2) -| (cost);
        \draw[decorate,decoration={brace, amplitude=3mm}] (-3.2, -1) -- node [below, yshift=-5mm, text width=1cm] {Ambient\\Space} (-5, -1);
        \draw[decorate,decoration={brace, amplitude=3mm}] (-0.5, -1) -- node [below, yshift=-5mm, text width=1cm] {Latent\\Space} (-3, -1);
        \draw[decorate,decoration={brace, amplitude=3mm}] (1.7, -1) -- node [below, yshift=-5mm, text width=1cm] {Ambient\\Space} (-0.3, -1);
    \end{tikzpicture}
\caption{Structure of a generic autoencoder.} 
\label{fig:autoencoder}
\end{figure}

Fig.~\ref{fig:autoencoder} depicts a generic autoencoder such as the one studied in~\cite{huang2019charting_representation_constrained}. 
A first DNN takes the input data from the ambient space to the latent space, by mapping a sample $\bm x_i$ to $\bm y_i = f_{\bm \theta}(\bm x_i)$.
Then, a second DNN\footnote{Here we assume without loss of generality that $\bm \theta$ represents the weights of both DNNs involved in the autoencoder.} takes the data from the latent space back to the ambient space, as $\bm{\hat x}_i = f'_{\bm \theta}(\bm y_i)$.
The cost function associated with autoencoder training is based on the Euclidean distance between the input and the output: 
\begin{equation}
\label{eq:costfct_autoencoder}
    C_a \equiv \frac{1}{|\mathcal S|}\sum_{i\in \mathcal S} \|\bm x_i - \bm{\hat x}_i\|^2.
\end{equation}
Note that it does not explicitly promote the equality in~\eqref{eq:distance_equality} between distances in the ambient and latent spaces.
\begin{figure}[ht!]
    \centering
    \scriptsize
    \begin{tikzpicture}[
        trapz/.style = { trapezium, draw, shape border rotate=180, minimum height=1cm, align=center},
        lab/.style = { text width=1.3cm, text centered },
        minus/.style = { circle, draw },
        box/.style = { draw, rounded corners, text centered }
    ]
        \node (training) at (0, 2.8) {Input data};
        \node[lab] at (-1, 1.5) (S1) {Sample 1};
        \node[lab] at (1, 1.5) (S2) {Sample 2};
        \coordinate (N)  at (0, 2.5);
        \draw[->] (N) -- (S1);
        \draw[->] (N) -- (S2);
        \node[trapz] at (-1, 0) (NN-S1) {DNN\\$f_{\bm \theta}$};
        \node[trapz] at (1, 0) (NN-S2) {DNN\\$f_{\bm \theta}$};
        \draw[->] (S1) -- (NN-S1) node[near start, left] {$\bm x_i$};
        \draw[->] (S2) -- (NN-S2) node[near start, right] {$\bm x_j$};
        \draw[decorate,decoration={brace, amplitude=3mm}] (2.2, 2.8) -- node [right, xshift=5mm, text width=1cm] {Ambient\\Space} (2.2, 0.9);
        \node[box, text width=1.8cm, fill=gray!20] at (3.5, 0.8) (weights) {Common\\Weights $\bm \theta$};
        \draw[->, help lines] (weights) -| (NN-S1.60);
        \draw[->, help lines] (weights) -| (NN-S2.60);
        \node[box, text width=1.8cm, minimum height=0.8cm, below=4 of N] (cost) {Cost function\\$C_s$};
        \draw[->] (NN-S1) -- (cost) node[near start, left] {$\bm y_i$};
        \draw[->] (NN-S2) -- (cost) node[near start, right] {$\bm y_j$};
        \draw[->] (S1) -- ++(-1.1, 0) |- (cost.west);
        \draw[->] (S2) -- ++(1.1, 0) |- (cost.east);
        \draw[decorate,decoration={brace, amplitude=3mm}] (2.2, 0) -- node [right, xshift=5mm, text width=1cm] {Latent\\Space} (2.2, -2);
    \end{tikzpicture}
\caption{Structure of a generic Siamese networks architecture. Both DNN block are identical, and use the same weights.}
\label{fig:siamese}
\end{figure}
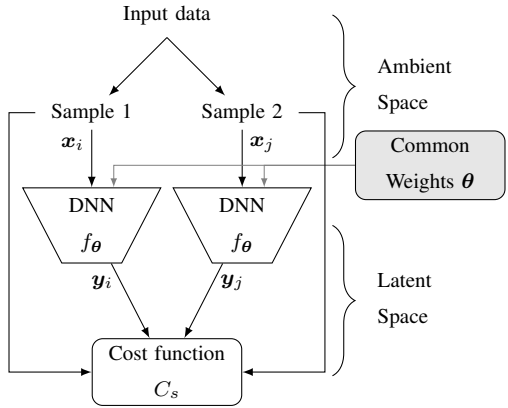

Fig.~\ref{fig:siamese} depicts a generic Siamese networks architecture as studied in~\cite{lei2019siamese} that uses a DNN to process a pair of input samples $\bm x_i$ and $\bm x_j$, yielding the respective images by $f_{\bm \theta}$ in the latent space denoted by $\bm y_i$ and $\bm y_j$.
The Siamese networks are trained to try and replicate the ambient space distance between the samples in the latent space, i.e., by minimizing 
\begin{equation}
\label{eq:costfct_siamese}
    C_s \equiv \frac{1}{|\mathcal T|} \sum_{(i, j) \in \mathcal T} \big(\|\bm x_i - \bm x_j\| - \|\bm y_i - \bm y_j\| \big)^2.
\end{equation}

Observe at this point that both cost functions \eqref{eq:costfct_autoencoder} and \eqref{eq:costfct_siamese} involve Euclidean distances in the ambient space.
However, it is notoriously difficult to define a meaningful distance between two vectors $\bm x_i$ and $\bm x_j$ having high dimensions.
Even in tightly controlled experimental set-ups, the Euclidean distance  $\|\bm x_i - \bm x_j\|$ will be high for practically all non-identical samples, due to the \emph{curse of dimensionality} phenomenon~\cite{Beyer1999}, as well as the acquisition impairments described before if one directly considers the CSI samples as the input set $\{ \bm x_i \}$ with no preprocessing.
This is a serious issue in practice, since the objective of preserving distances between $\mathcal X$ to $\mathcal Y$ is pointless if the distance on $\mathcal X$ does not make sense in the first place.

\section{Triplet-based Dimensionality Reduction}
\label{sec:triplets}
To reduce the aforementioned effect, most of the current state-of-the-art approaches to channel charting try to preprocess the CSI data into a form where the Euclidean distance is meaningful.
In this paper, we adopt a preprocessing step that aims at removing the corruption seen in real-world wireless deployments, and we propose a new approach capable of simultaneously learning a meaningful distance on $\mathcal X$ and performing dimensionality reduction.
The preprocessing has been extensively described in~\cite{Localization_Globecom} and is reproduced in Section~\ref{sec:feature_transformation}; the network structure is described in the current section.
In the considered DNN-based approach, the definition of the distance $d(\cdot, \cdot)$ is based on the parametric mapping in~\eqref{DNN_mapping}:
\begin{equation}
  d_{\bm \theta}(\bm x_i, \bm x_j) = \left\| f_{\bm \theta}( \bm x_i ) -f_{\bm \theta} ( \bm x_j ) \right\|
\end{equation}
where $f_{\bm \theta}$ is the DNN transfer function.
Note that, conversely to the approaches discussed in Sections~\ref{sec:problem_description} and \ref{sec:sota_charting}, the above definition completely avoids any reliance on the Euclidean distance on $\mathcal X$.
We now present a method to learn a projection from $\mathbb R^D$ to $\mathbb R^d$ (by estimating the optimal $\bm \theta$) that makes the above distance meaningful.

\subsection{Distance Learning}
\begin{figure}[ht!]
  \centering
  \scriptsize
  \begin{tikzpicture}[
      trapz/.style = { trapezium, draw, shape border rotate=180, minimum height=1cm, align=center},
      lab/.style = { text width=1cm, text centered },
      minus/.style = { circle, draw },
      box/.style = { draw, rounded corners, text centered }
  ]
      \node (training) at (0, 2.8) {Input data};
      \node[lab] at (-2, 1.7) (close) {Close\\sample};
      \node[lab] at (0, 1.5) (anchor) {Anchor};
      \node[lab] at (2, 1.7) (far) {Far\\sample};
      \coordinate (N)  at (0, 2.5);
      \draw[->] (N) -- (anchor);
      \draw[->] (N) -- (close);
      \draw[->] (N) -- (far);
      \node[trapz] at (-2, 0) (NN-close) {DNN\\$f_{\bm \theta}$};
      \node[trapz] at (0, 0) (NN-anchor) {DNN\\$f_{\bm \theta}$};
      \node[trapz] at (2, 0) (NN-far) {DNN\\$f_{\bm \theta}$};
      \draw[->] (anchor) -- (NN-anchor) node[near start, left] {$\bm x_i$};
      \draw[->] (close) -- (NN-close) node[near start, left] {$\bm x_j$};
      \draw[->] (far) -- (NN-far) node[near start, right] {$\bm x_k$};
      \draw[decorate,decoration={brace, amplitude=3mm}] (2.8, 2.8) -- node [right, xshift=5mm, text width=1cm] {Ambient\\Space} (2.8, 0.9);
      \node[box, text width=1.8cm, fill=gray!20] at (4.1, 0.8) (weights) {Common\\Weights $\bm \theta$};
      \draw[->, help lines] (weights) -| (NN-anchor.60);
      \draw[->, help lines] (weights) -| (NN-far.60);
      \draw[->, help lines] (weights) -| (NN-close.60);
      \node[box, text width=1.8cm, minimum height=0.8cm, below=1 of NN-anchor] (cost) {Cost function\\$C_t$};
      \draw[->] (NN-close) -- (cost) node[near start, left] {$\bm y_j$};
      \draw[->] (NN-anchor) -- (cost) node[near start, left=0.1] {$\bm y_i$};
      \draw[->] (NN-far) -- (cost)  node[near start, right=0.1] {$\bm y_k$};
      \draw[decorate,decoration={brace, amplitude=3mm}] (2.8, 0) -- node [right, xshift=5mm, text width=1cm] {Latent\\Space} (2.8, -2);
  \end{tikzpicture}
\caption{Triplet network architecture as introduced in \cite{Schroff2015} as ``FaceNet''. The three DNN blocks are identical, and use common weights.}
\label{fig_triplet}
\end{figure}
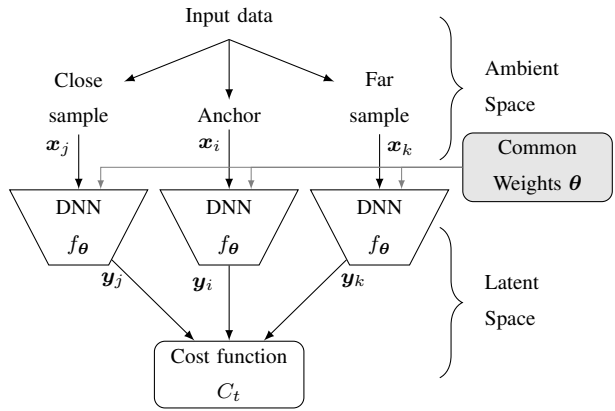

Assume that training data is available in the form of triplets of samples $(\bm x_i,\bm x_j,\bm x_k)$ with $(i, j, k) \in \mathcal T \subset \mathcal S^3$, for which it is known that the respective distances of $\bm x_i$ to $\bm x_j$ and $\bm x_k$ fulfill
\begin{equation}  \label{eq_inequality_x}
d(\bm x_i, \bm x_j) \leq d(\bm x_i, \bm x_k)
\end{equation}
i.e. $\bm x_j$ is closer than $\bm x_k$ to a reference sample $\bm x_i$.
In the literature, $\bm x_j$ is sometimes called the \emph{positive} sample, whereas $\bm x_k$ is called the \emph{negative} sample.
Distance learning can be performed by training the DNN such that the inequality \eqref{eq_inequality_x} is promoted for the triplets in $\mathcal T$, as proposed e.g. in \cite{NIPS2009_3658_BoostMetric,Schroff2015}.
Naively, this could be achieved by minimizing the sample mean over a training dataset of the difference between the two terms of the inequality \eqref{eq_inequality_x}, defined as
\begin{multline}
C_t = \frac{1}{|\mathcal T|} \sum_{(i, j, k) \in \mathcal T} \Big( \left\| f_{\bm \theta}( \bm x_i ) -f_{\bm \theta} ( \bm x_j ) \right\| \\
 - \left\| f_{\bm \theta}( \bm x_i ) -f_{\bm \theta} ( \bm x_k ) \right\| \Big)^+ \label{eq_triplets} 
\end{multline}
where $(x)^+$ is a shorthand for the function $\max(x, 0)$.
The solution of the optimization problem 
\begin{equation}
\bm \theta^* = \arg\min_{\bm \theta} C_t
\end{equation}
defines the learned distance, which is then taken as $d=d_{\bm \theta^*}$.
This triplet-based joint distance learning and dimensionality reduction process is summarized in Fig.~\ref{fig_triplet}.

Note that the cost function in \eqref{eq_triplets} has a trivial solution, which maps all samples to the same point in $\mathcal Y$ and achieves zero cost.
In order to avoid this uninteresting solution, some slack can be added to the cost function through a constant $M$ so that only the points for which $\left\| f_{\bm \theta}( \bm x_i ) -f_{\bm \theta} ( \bm x_j ) \right\| + M > \left\| f_{\bm \theta}( \bm x_i ) -f_{\bm \theta} ( \bm x_k ) \right\|$ contribute to the cost, yielding the \emph{margin cost}
\begin{equation}  \label{eq_triplets_margin}
  \frac{1}{|\mathcal T|}\sum_{(i, j, k) \in \mathcal T}  \Big(d_{\bm \theta}(\bm x_i, \bm x_j) - d_{\bm \theta}(\bm x_i, \bm x_k) + M\Big)^+.
\end{equation}
The margin encourages the distance between the anchor and far point to be larger than the distance between the anchor and close point.
By a proper selection of the triplets, we expect this property to be true with high probability.
The margin can be chosen as $M=1$ without loss of generality (a different margin would be equivalent up to some global scaling of the chart coordinates).
An alternative to the margin cost consists in a logarithmic averaging of the per-sample cost, where large differences are more penalized than small ones, leading to the \emph{exponential cost}
\begin{equation}  \label{eq_triplets_exp}
  \frac{1}{|\mathcal T|}\log \Big(\sum_{(i, j, k) \in \mathcal T}  \exp \left(d_{\bm \theta}(\bm x_i, \bm x_j) - d_{\bm \theta}(\bm x_i, \bm x_k) \right)\Big).
\end{equation}
The cost function in~\eqref{eq_triplets_exp} can also be viewed as a soft-maximum of the distance differences taken among the training set.
Taking the logarithm reduces the dynamic range of the objective, and brings back the value in a similar scale as the distances themselves.
This notion of scale becomes important when moving towards e.g. semi-supervised training which is discussed in Section~\ref{sec:semi_supervised_training}.

\subsection{Time-Based Training Dataset Construction}
\label{sec:training_qualitative_information}
The fact that the triplet-based approach outlined above enables distance learning based on inequalities can be advantageous in scenarios where the definition of an appropriate distance on the ambient space is elusive.
When considering CSI data, the fact that \eqref{eq_inequality_x} involves an inequality (which is admittedly a less informative statement than an equality constraint) is beneficial since it allows to make use of qualitative information for training.
In particular, one can make such a statement based on the \emph{sampling epoch}, i.e. the exact time at which the CSI sample was measured; indeed, due to the time-correlated nature of CSI, it is reasonable to expect that two CSI samples acquired for the same user within a short time interval from each other (during which little or no change in the environment is expected to take place) are, at least with high probability, \emph{more similar} than two samples selected randomly in the dataset, allowing to construct a set $\mathcal T$ for which~\eqref{eq_inequality_x} holds with high probability.
This set can be subsequently used for distance learning as in~\eqref{eq_triplets}.

\begin{figure}[ht!]
  \centering
  \scriptsize
  \begin{tikzpicture}[>=latex, transform shape, scale=0.94]
    \fill[col5!20] (-1.8, -0.2) rectangle (1.8, 0.2);
    \fill[col2!20] (-4, -0.2) rectangle (-1.8, 0.2);
    \fill[col2!20] (1.8, -0.2) rectangle (4, 0.2);
    \draw[help lines, ->] (-4.2, 0) -- (4.5, 0) node [right] {Time};
    \draw (0, 0) -- ++(0, 0.5) node [black, fill=white] {$t_i$};
    \node at (0, 0.8) {Anchor};
    \draw[fill=black] (0, 0) circle (0.05);
    \draw (1.3, 0) -- ++(0, 0.5) node [black, fill=white] {$t_j$};
    \node at (1.3, 0.77) {Close sample};
    \draw[fill=black] (1.3, 0) circle (0.05);
    \draw (3.5, 0) -- ++(0, 0.5) node [black, fill=white] {$t_k$};
    \node at (3.5, 0.77) {Far sample};
    \draw[fill=black] (3.5, 0) circle (0.05);
    \node at (-1.8, -0.6) {$t_i-T_c$};
    \node at (1.8, -0.6) {$t_i+T_c$};
    \node at (-4, -0.6) {$t_i-T_f$};
    \node at (4, -0.6) {$t_i+T_f$};
  \end{tikzpicture}
  \caption{Illustration of the triplet selection approach based on the sampling epoch (see \eqref{eq:triplet_time_threshold}).}
  \label{fig:curriculum_thresholds}
\end{figure}
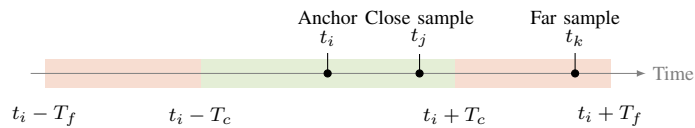

Assume that timestamps indicating the absolute time when a CSI sample was acquired are available for the whole dataset, and let $t_i$ denote the timestamp associated with the preprocessed CSI sample $\bm x_i$, $i \in \mathcal N$.
The main parameter in this approach consists in an inner threshold $T_c$ on the time span over which the channel is expected to show significant self-similarity (this is often denoted as the channel coherence time).
For generality and in order to support further extensions such as the curriculum learning in Section~\ref{sec:curriculum}, an outer threshold $T_f$ is also considered.
Consequently, the set $\mathcal T$ of training indices is defined as
\begin{equation}
  \mathcal T \subset \left\{(i, j, k) \in \mathcal S^3 \quad \text{ s.t. } \quad
  \begin{aligned}
    0 &< |t_j - t_i| \leq T_c \\ 
    T_c &< |t_k - t_i| \leq T_f 
  \end{aligned}\right\}
  \label{eq:triplet_time_threshold}
\end{equation}
These thresholds are illustrated in Fig.~\ref{fig:curriculum_thresholds}.

\section{Dimensionality Reduction Quality Metrics}
\label{sec:measuring_chart_quality}
In principle, any measure of the quality of the obtained chart should be closely linked to the objective function used to obtain the chart.
For most quality metrics, the distances are compared between the latent space and a \emph{reference} space.
The reference space in many applications is taken as the ambient space, since properties of this ambient space are expected to be transferred to the latent space through the dimensionality reduction algorithm.
Here, we propose also to consider the user location in the \emph{geographical} space as the reference space, for all quality metrics involving distances.
The rationale is the same as for the dimensionality reduction algorithm itself: quality metrics relying on distances in the ambient space are worthless when the available distances on the ambient space are unreliable.
For applications of channel charting where the chart is intended to act as a surrogate radio map, checking properties with respect to to the geographical space is relevant.
We will therefore assume that additional user location information is available for the evaluation of the chart quality metric only (while charting itself remains self-supervised).
In the following, the latent space samples $\{\bm y_i\}$ are assumed computed and known.
Denote as $\{\bm r_i\}$ the matching samples in the reference space.
Depending on the context, $\{\bm r_i\} = \{\bm x_i\}$ if the reference is the ambient space, and $\{\bm r_i\} = \{\bm p_i\}$ if the reference is the geographical space, where $\{\bm p_i\}$ denotes the set of geographical positions.
Besides the excess variance defined in Section~\ref{sec:excess_variance} and the optimal scaling, rotation, and/or reflection of Section~\ref{sec:scale_and_rotate}, these metrics have been introduced and used in e.g.~\cite{Studer_etal_charting,huang2019charting_representation_constrained}, and subsequent works.
We reproduce them here for completeness.

The increasing number of metrics in the channel charting literature seems to indicate that measuring the quality of a chart in absolute terms is in fact a difficult problem.
However, it is possible to give some hint about what they do measure about the chart, and which application they may be more suited for:
\begin{itemize}
  \item Trustworthiness and continuity being based on neighborhood information are therefore amenable to application regarding clustering.
  \item The excess variance measures the ``noisiness'' of trajectories and thus indicates how well one could predict movement on the channel chart.
  \item The Kruskal stress and the optimal scaling, rotation, and/or reflection metric try to capture whether the global and local geometries of the chart respect that of the reference space.
\end{itemize}

\subsection{Trustworthiness and Continuity}

A common way to assess the quality of the chart relates to the preservation of neighborhoods using a metric known as \emph{trustworthiness}.
Fix a given neighborhood size $K$, and let $\mathcal N_i^K \subset \mathcal S$ be the set of $K$ neighbors of the point indexed by $i$ in the reference space.
We order this set by the distance between a point indexed by $u \in \mathcal N_i^K$ and the point indexed by $i$.
The rank of the point indexed by $u$ among the ordered $K$-neighbors is $n_i(u)$.
We define similarly $\hat{\mathcal N}_i^K \subset \mathcal S$ and $\hat n_i(u)$ for the latent space, so that the former is the $K$-neighborhood of the point indexed by $i$ in this space, and the latter is the rank of the point indexed by $u$ among all $K$-neighbors.
Trustworthiness is defined as
\begin{equation}
  \mathsf{TW} =  1 - \frac{2}{C \left| \mathcal S \right|}\sum_{i \in \mathcal S}\sum_{u \in \hat{\mathcal N}_i^K} \big(n_i(u) - K\big)^+
  \label{eq:trustworthiness}
\end{equation}
where $C = K(2N - 3K - 1) / 2$ is a normalization constant.
Trustworthiness is a measure of neighborhoods, and can be defined over different distances for the sets $\mathcal N_i^K$ and $\hat{\mathcal N}_i^K$.
The trustworthiness formula penalizes samples that are neighbors in the latent space but not in the reference space.
A related metric, \emph{continuity}, is defined similarly by swapping the roles of the reference and latent spaces:
\begin{equation}
    \mathsf{CT} = 1 - \frac{1}{NC} \sum_{i \in \mathcal S} \sum_{u \in \mathcal N_i^K} \big(\hat n_i(u) - K\big)^+
    \label{eq:continuity}
\end{equation}
Both $\mathsf{TW}$ and $\mathsf{CT}$ are equal to 1 when all samples have identical sets of neighbors in the reference and latent spaces.
High trustworthiness and continuity values are thus an indication that sample neighborhoods are similar between the reference space and the embedding.

\subsection{Optimal Scaling, Rotation, and/or Reflection}
\label{sec:scale_and_rotate}
Trustworthiness and continuity are local metrics: they measure how local properties (neighborhoods) are preserved between the reference space and the latent space.
A more global similarity metric consists in the optimal scaling, rotation, and/or reflection (SR) metric, defined as the residual distance between the original points and the corresponding embedding points after the aforementioned operations to match the points have been performed.
Let us define the sample means
\begin{align}
  \bm{\bar y} = \frac 1N \sum_{i \in \mathcal S} \bm y_i \qquad &\text{and} \qquad \bm{\bar r} = \frac 1N \sum_{i \in \mathcal S} \bm r_i \\
\intertext{so that we can center and rescale $\bm y_i$, and center $\bm r_i$ as follows}
  \label{eq:rescale_latent_wrt_reference}
  \bm{\hat y}_i = \frac{\sigma_r}{\sigma_y} \bigg( \bm y_i -\bm{\bar y} \bigg) \qquad &\text{and} \qquad \bm{\hat r_i} = \bm r_i - \bm{\bar r}
\end{align}
with
\begin{equation*}
  \sigma_r = \sqrt{\frac 1N \sum_{i \in \mathcal S} \| \bm r_i - \bm{\bar r} \|^2}, \qquad \sigma_y = \sqrt{\frac 1N \sum_{i \in \mathcal S} \| \bm y_i - \bm{\bar y} \|^2}.
\end{equation*}
Assuming that all the $\bm{\hat r}_i$ (respectively $\bm{\hat y}_i$) are organized as the rows of matrix $\bm{\hat R}$ (respectively $\bm{\hat Y}$), this is known as the Procrustes problem and a result by Schönemann~\cite{Schonemann1966} allows to compute the orthogonal matrix $\bm W$ that bests maps $\bm{\hat Y}$ to $\bm{\hat R}$, i.e. find 
\begin{equation}
  \bm W = \arg \min_{\bm \Omega} \bm \| \bm{\hat Y} \bm \Omega - \bm{\hat R} \|_F \qquad \bm \Omega^T \bm \Omega = \bm I.
\end{equation}
The optimal $\bm W$ is obtained as $\bm W = \bm U \bm V^T$, where $\bm U \bm S \bm V^T$ is the singular value decomposition of $\bm{\hat Y}^T \bm{\hat R}$.
The $\mathsf{SR}$ metric is defined as the average squared distance between the points in the transformed set and the centered reference set:
\begin{equation}
  \label{eq:optimal_sr}
  \mathsf{SR} = \frac 1N \sum_{i \in \mathcal S} \| \bm{\hat r}_i - \bm{\tilde y}_i \|^2
\end{equation}
where $\bm{\tilde y}_i$ denotes the $i$-th row of  $\bm{\tilde Y} = \bm{\hat Y} \bm W$.
Note that this indeed gives a high-level view of the quality of the chart but does not allow local discrepancies.
The metric can be made more local by applying a different scaling depending on the density of the points in the area~\cite{deSilva2002}.

\subsection{Kruskal Stress}

Kruskal stress was originally designed as an optimization cost for general dimensionality reduction applications~\cite{Kruskal1964}.
In essence, it measures how well distances in the latent space reflect distances in the reference space globally.
It is defined as follows:
\begin{equation}
    \mathsf{KS} = \sqrt{\frac{\sum_{i \in \mathcal S} \sum_{j \in \mathcal S \backslash i} (\|\bm r_j - \bm r_i\| - \beta\|\bm y_j - \bm y_i\|)^2}{\sum_{i \in \mathcal S} \sum_{j \in \mathcal S \backslash i} \|\bm r_j - \bm r_i\|^2}}.
\end{equation}
Lower Kruskal stress values indicate a better fit between the pairwise distances in the two spaces.
When used as a cost function, the scale of the latent space should be the same as the reference space since any difference in scale would be detrimental to the cost.
We adopt here the scaling of~\cite{huang2019charting_representation_constrained} and choose
\begin{equation}
  \beta = \frac{\sum_{i \in \mathcal S} \sum_{j \in \mathcal S \backslash i} \|\bm r_j - \bm r_i\|\|\bm y_j - \bm y_i\|}{\sum_{i \in \mathcal S} \sum_{j \in \mathcal S \backslash i} \|\bm y_j - \bm y_i\|^2}
\end{equation}
which ensures that the resulting metric is below 1.

\subsection{Excess Variance}
\label{sec:excess_variance}
Finally, we include a variance measure, which echoes the box-counting measures and intrinsic dimensionality of~\cite{Camastra2002}.
Start by scaling globally the latent space with respect to the reference space, as in \eqref{eq:rescale_latent_wrt_reference}.
The excess variance with respect to the reference for a neighborhood of size $K$ is defined as
\begin{equation}
  \mathsf{EV} = \label{eq:excess_variance}
  \frac 1{KN} \sum_{i \in \mathcal S} \bigg( \sum_{u \in \hat{\mathcal N}_i^K} \|\bm{\hat y}_u - \bm {\bar \mu}_i^K\|^2 - \sum_{u \in \mathcal N_i^K} \|\bm{\hat r}_u - \bm {\mu}_i^K\|^2\bigg) 
\end{equation}
where $\bm{\bar \mu}_i^K$ is the average position of the $K$-neighbors of $\bm{\hat y}_i$ in the latent space, and $\bm {\mu}_i^K$ is the average position of the $K$-neighbors of $\bm{\hat r}_i$ in the reference space.

\section{Experimental Results}
\label{sec:experimental_results}
Experimental validation of the proposed triplet-based approach was performed using data collected on a Huawei commercial 4G system (BS and smartphone).
The BS used for these measurements is equipped with 64 antennas arranged regularly in a rectangular array of 4 $\times$ 8 and each antenna element has 2 polarizations.
It it located on a rooftop, and collects uplink CSI samples at a frequency of 200 Hz, leading to a sampling period of 5 ms, together with a timestamp.
Each sample consists in 288 subcarriers spaced over a 10 MHz bandwidth.
The UE was held by a human moving outdoors at pedestrian speed, for which a Global Navigation Satellite System (GNSS) position was also recorded.
We provide a short summary of the parameters in Table~\ref{tab:summary}.
The same data was also used in the context of the localization experiments presented in \cite{Localization_Globecom}.
\begin{table}[hb]
  \centering
  \caption{Main characteristics of the experiment used to gather the datasets used in this article.}
  \label{tab:summary}
  \begin{tabular}{lc}
    \toprule
    Number of antennas & 32 dual-polarized \\
    Antenna array shape & Rectangular \\
    Repartition (azimuth/zenith) & 8/4 \\
    Frequency bandwidth & 10 MHz \\
    Carrier frequency & 2.5 GHz \\
    Number of subcarriers & 288, equispaced \\
    Sampling period & 5 ms (200 Hz) \\
    UE velocity & Pedestrian, around 1 m/s \\
    \bottomrule
  \end{tabular}
\end{table}

\subsection{Data preprocessing}
\label{sec:feature_transformation}
In our previous work~\cite{Localization_Globecom}, we developed a preprocessing step for CSI data that aims at removing the phase corruption seen in practical measurements.
This preprocessing step has been shown to work adequately in geolocation application in wireless network.
For completeness, we briefly describe this transformation here, and we refer the reader to the original article for additional rationale and discussion regarding this process.
Let $i$, $f$, $n$, $m$ and $p$ index the time, frequency, vertical antenna, horizontal antenna, and polarization; the CSI sample at time $i$ is denoted by $h_i(p, m, n, f)$.
In practical cellular systems, a number of hardware and protocol design characteristics might corrupt the wireless channel state as it is seen by the BS.
Dominant impairements are usually related to manufacturing variations within the radio-frequency components connected to each antenna element, clock and frequency offsets between the UE and the BS, and timing advance variations at the BS.
These impairments mostly affect the phase of the channel samples on each subcarrier and antenna.
On the other hand, relevant information in the channel lies in the phase differences between the samples obtained in e.g. neighboring antennas or subcarriers.
In order to remove impairments while keeping as much information in the data as possible, it was preprocessed as follows:
\begin{enumerate}
  \item Apply a 2D Fourier transform to the channel measurements $h_i(p, m, n, f)$ to form the beam domain channel measurements $\tilde{h}_i(p, z, a, f)$ where $z$ and $a$ index the zenith and azimuth beam respectively. Such an approach has been shown to be beneficial in~\cite{Studer_etal_charting}.
  \item Compute the absolute value of the frequency-domain complex autocorrelation $$c_i(p, z, a, \delta) = \mathbb E\Big[\tilde{h}_i(p, z, a, f)\tilde{h}^*_i(p, z, a, f + \delta)\Big]$$ where $\delta$ denotes the autocorrelation lag in the frequency dimension.
  The expectation is taken over the frequency dimension. Further downsample and truncate the autocorrelation in the $\delta$ dimension, and keep 16 equally spaced elements so that $\delta \in \{0, 4, 8, \cdots, 60\}$.
  \item Take the logarithm of the result, i.e. $\log |c_i(p, z, a, \delta)|$.
\end{enumerate}
These steps yields input data samples consisting of 1024 real coefficients, down from 18,432 complex coefficients in the raw samples.

\subsection{Neural Network Building Blocks}
\label{sec:nn_building_block}
\begin{figure}[ht!]
  \centering
  \includegraphics[width=0.6\columnwidth]{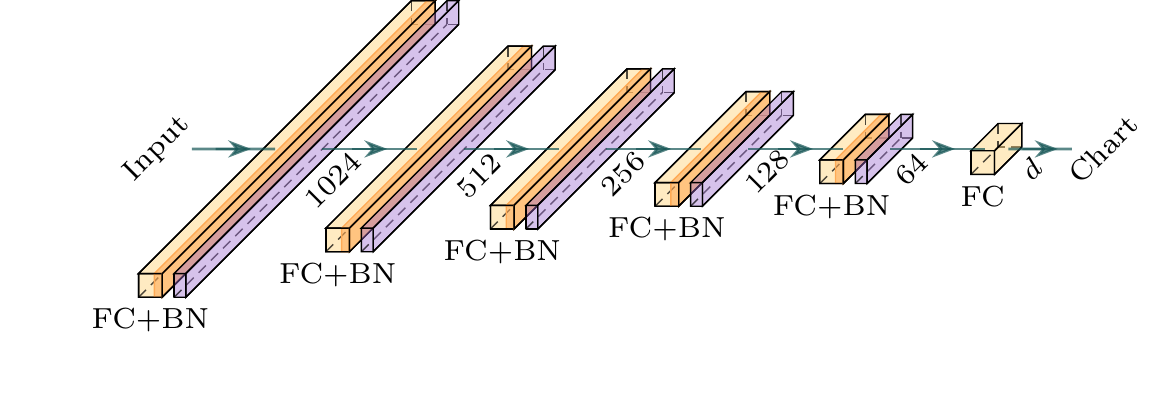}
  \caption{Neural network composed of 6 fully connected (FC) layers and 5 batch normalization (BN) layers. Each FC layer except the last one apply a ReLU nonlinearity.}
  \label{fig:localization_nn}
\end{figure}
The fully-connected neural network depicted in Fig.~\ref{fig:localization_nn}, represented by $f_{\bm \theta}$ in Sections~\ref{sec:sota_charting} and \ref{sec:triplets}, is used as a building block for all NN-based algorithms studied in this article: the symmetric autoencoder (Fig.~\ref{fig:autoencoder}) and the siamese network (Fig.~\ref{fig:siamese}) of Section~\ref{sec:sota_charting}, as well as the triplet-based channel charting (Fig.~\ref{fig_triplet}) of Section~\ref{sec:triplets}.
In the extensions of triplet-based channel charting in Section~\ref{sec:consecutive_samples}, successive samples are considered as an input to the network; this case includes a more complex network structure for comparison purposes, which will be described therein.

\begin{figure*}[t]
  \centering
  \begin{subfigure}[t]{0.27\textwidth}
      \includegraphics[width=\textwidth]{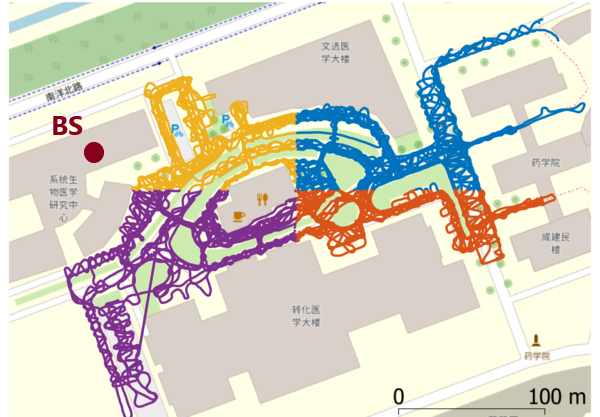}
      \caption{Geographic position}
      \label{fig:pedestrian_GNSS}
  \end{subfigure}%
  \qquad%
  \begin{subfigure}[t]{0.27\textwidth}
      \includegraphics[width=\textwidth]{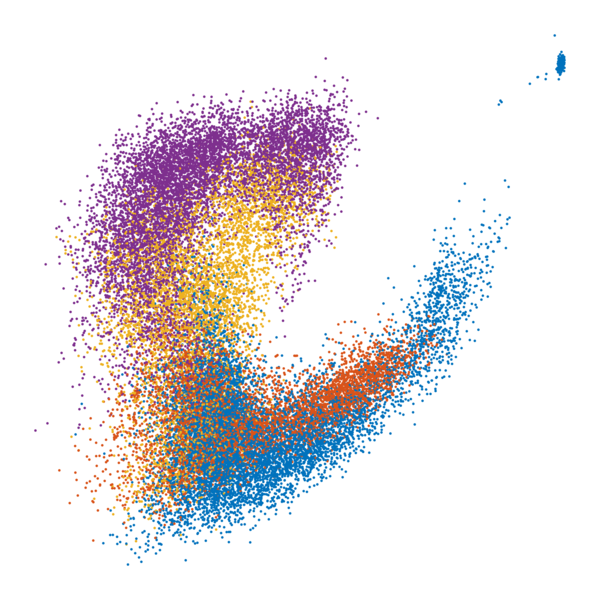}
      \caption{Principal component analysis (PCA)}
      \label{fig:pedestrian_PCA}
  \end{subfigure}%
  \qquad%
  \begin{subfigure}[t]{0.27\textwidth}
      \includegraphics[width=\textwidth]{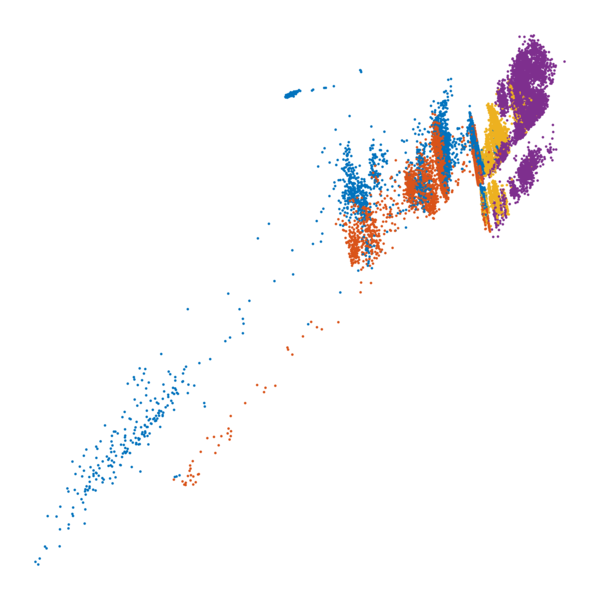}
      \caption{Autoencoder}
  \end{subfigure}
  \\
  \begin{subfigure}[t]{0.27\textwidth}
      \includegraphics[width=\textwidth]{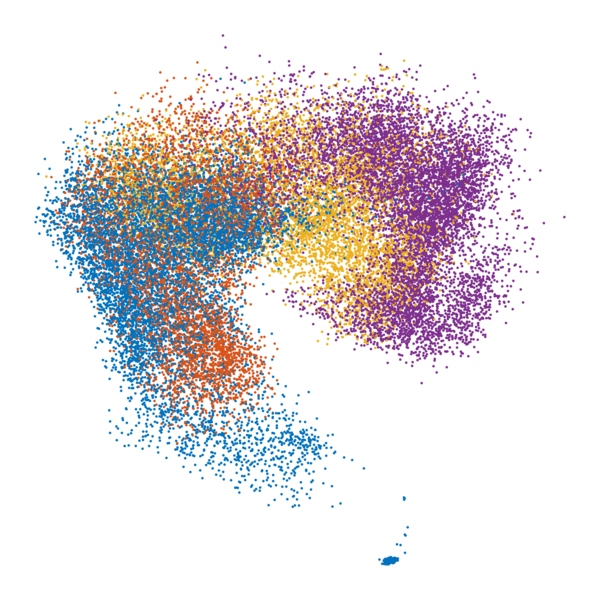}
      \caption{Siamese network}
  \end{subfigure}%
  \qquad%
  \begin{subfigure}[t]{0.27\textwidth}
      \includegraphics[width=\textwidth]{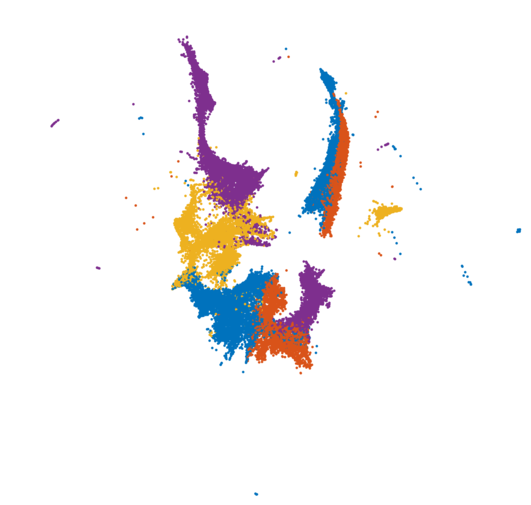}
      \caption{UMAP}
      \label{fig:pedestrian_UMAP}
  \end{subfigure}%
  \qquad%
  \begin{subfigure}[t]{0.27\textwidth}
      \includegraphics[width=\textwidth]{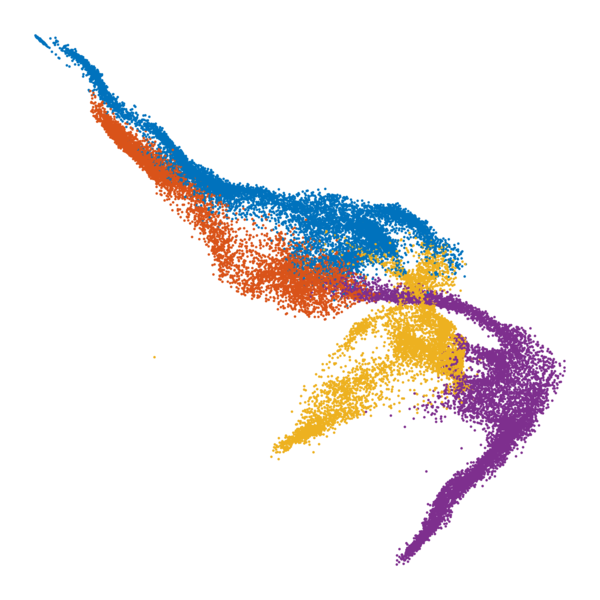}
      \caption{Triplets (margin cost)}
      \label{fig:pedestrian_triplet_margin}
  \end{subfigure}
  \caption{Visual representation of the different channel charts obtained on Dataset 1.}
  \label{fig:dataset_1_charts}
\end{figure*}

The number of parameters in the network of Fig.~\ref{fig:localization_nn} adds up to 1.7 million single precision floats.
In all cases, the network used a mean-squared loss and the Adam optimizer~\cite{Kingma2015}.
The implementation relies on Tensorflow~\cite{tensorflow2015}, with training over 100 epochs and a batch size of 1000.
For the Siamese and triplet approaches, 1000 batches of 2000 pairs or triplets of samples were respectively selected.
The runtime cost of training all the DNN based approach is similar: on a NVidia Tesla V100 GPU card with the aforementioned parameters it ranges between 20 minutes to 1 hour on a dataset containing 3 hours of data.
The main differences in runtime between the approaches lie in the time it takes to build the training batches.
The proposed triplet-based network in particular imposes more constraints on the training dataset than the auto-encoders or the Siamese networks.
UMAP takes an order of magnitude longer to produce results on such a big dataset, which is in the order of days on a dual Intel Xeon 8164 CPU with 104 computation cores.

\begin{figure*}[t]
  \centering
\begin{subfigure}[t]{0.27\textwidth}
    \includegraphics[width=\textwidth]{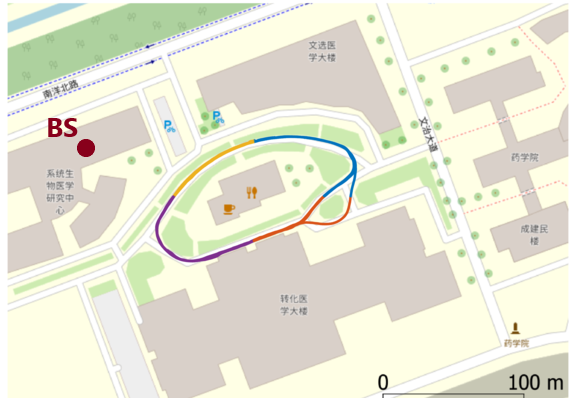}
    \caption{Geographic position}
    \label{fig:circles_GNSS}
\end{subfigure}%
\qquad%
\begin{subfigure}[t]{0.27\textwidth}
    \includegraphics[width=\textwidth]{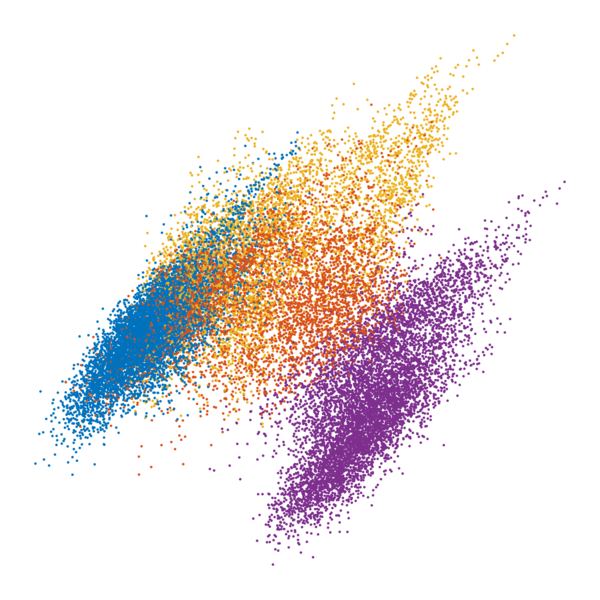}
    \caption{Principal Component Analysis (PCA)}
    \label{fig:circles_PCA}
\end{subfigure}%
\qquad%
\begin{subfigure}[t]{0.27\textwidth}
    \includegraphics[width=\textwidth]{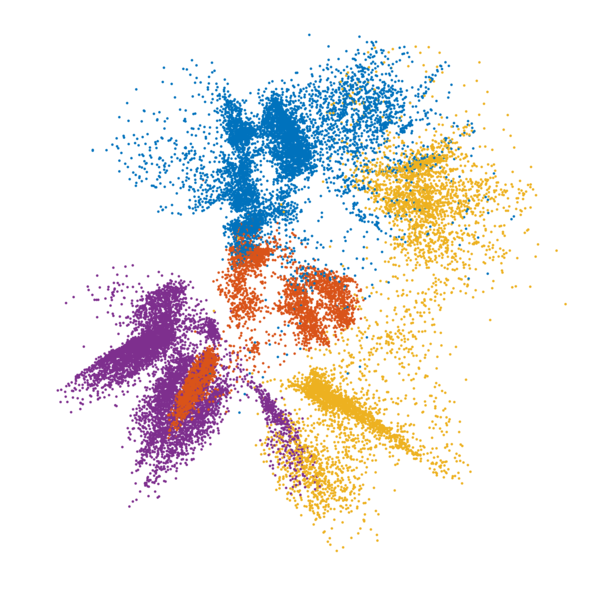}
    \caption{Autoencoder}
\end{subfigure}
\\
\begin{subfigure}[t]{0.27\textwidth}
    \includegraphics[width=\textwidth]{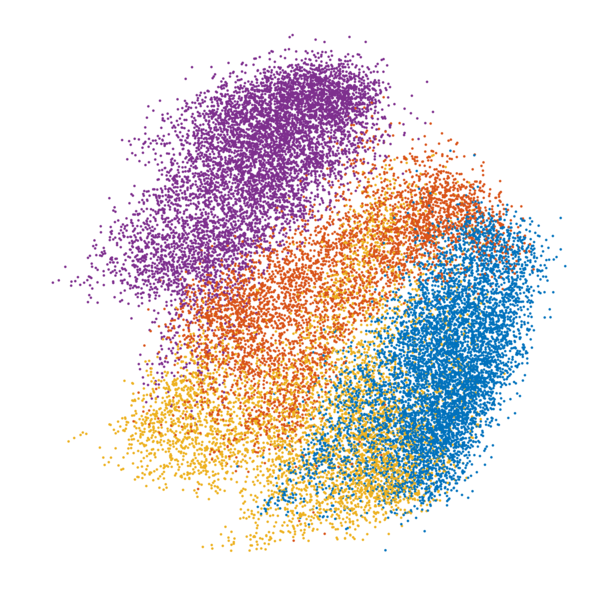}
    \caption{Siamese network}
\end{subfigure}%
\qquad%
\begin{subfigure}[t]{0.27\textwidth}
    \includegraphics[width=\textwidth]{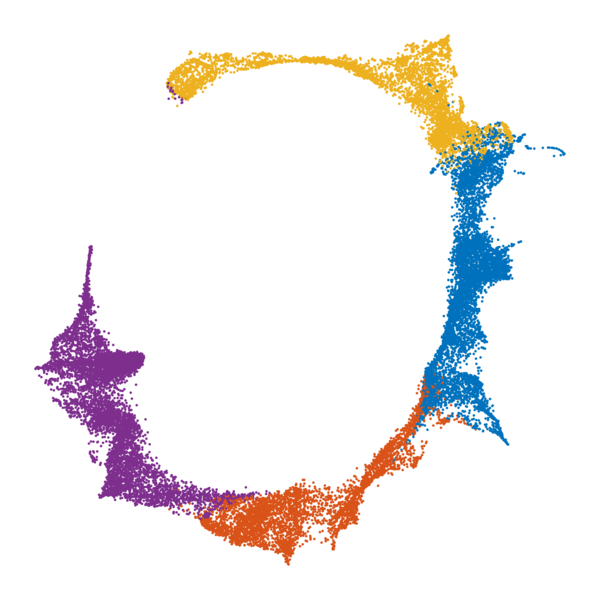}
    \caption{UMAP}
    \label{fig:circles_UMAP}
\end{subfigure}%
\qquad%
\begin{subfigure}[t]{0.27\textwidth}
    \includegraphics[width=\textwidth]{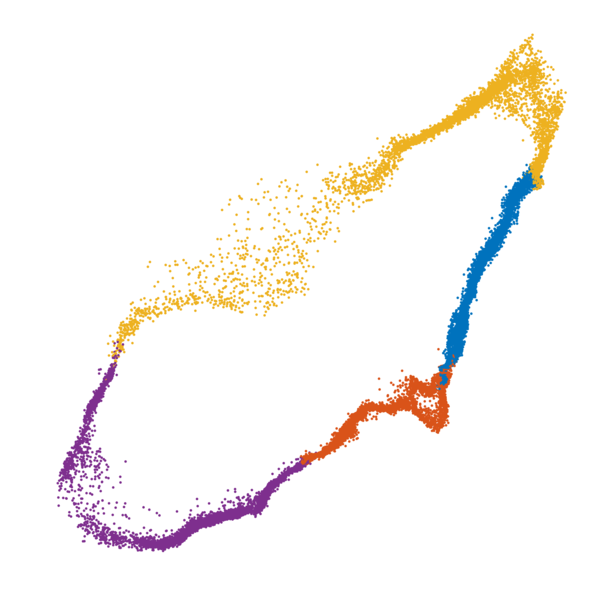}
    \caption{Triplets (margin cost)}
    \label{fig:circles_triplet_margin}
\end{subfigure}
\caption{Visual representation of the different channel charts obtained on Dataset 2. The colors are computed from the position to distinguish the 4 quadrants. The triplet network trained with the log-sum-exp cost has a similar shape to the one trained with the margin cost.}
  \label{fig:dataset_2_charts}
\end{figure*}

\subsection{Charting Results}

We apply all the charting methods considered in this article on 2 datasets that we describe hereafter.
Besides the DNN-based methods detailed in Section~\ref{sec:sota_charting}, a number of ``classical'' dimensionality reduction approaches were also implemented.
Specifically, we evaluated the UMAP algorithm~\cite{McInnes2018}, an algorithm of the Isomap family with good performance on our data and moderate complexity.
All algorithms were used to produce charts of dimension $d=2$.

\subsubsection{Dataset 1}
The first considered dataset contains 3 hours of data (approximately 2.1 million samples) gathered over 2 different days using different handsets carried by human experimenters walking at pedestrian speed.
The GNSS position of the handset is shown in Fig.~\ref{fig:pedestrian_GNSS}, while figures~\ref{fig:pedestrian_PCA} through \ref{fig:pedestrian_triplet_margin} depict the charts created using different approaches.
In all figures, the samples have been assigned arbitrary colors according to the the quadrants of the geographic positions (see Fig.~\ref{fig:pedestrian_GNSS}), in order to enable comparison between the charts and with the GNSS position data.
The triplet network trained with the log-sum-exp cost yields a visually similar result to the one trained with the margin cost, and therefore results for this case are not depicted.

The value of the similarity metrics discussed in Section~\ref{sec:measuring_chart_quality}, namely Kruskal stress ($\mathsf{KS}$), trustworthiness ($\mathsf{TW}$), continuity ($\mathsf{CT}$), optimal scaling and rotation ($\mathsf{SR}$) and excess variance ($\mathsf{EV}$), are shown in Table~\ref{tab:pedestrian}.
Note that 0 is best for the $\mathsf{SR}$, $\mathsf{KS}$ and $\mathsf{EV}$ metrics, while 1 is best for $\mathsf{TW}$ and $\mathsf{CT}$.

\begin{table}[hb!]
  \caption{Metric comparisons for different charting methods applied to Dataset 1.}
  \label{tab:pedestrian}
  \centering
  \begin{tabular}{rrrrrr}
    \toprule
      Method & \multicolumn{1}{c}{$\mathsf{KS}$} & \multicolumn{1}{c}{$\mathsf{TW}$} & \multicolumn{1}{c}{$\mathsf{CT}$} & \multicolumn{1}{c}{$\mathsf{SR}$} & \multicolumn{1}{c}{$\mathsf{EV}$} \tabularnewline
    \midrule
    MDS & 0.524 & 0.843 & 0.894 & 79.2 & 0.08 \tabularnewline 
    UMAP & 0.469 & 0.951 & 0.923 & 87.7 & 0.22 \tabularnewline 
    Autoencoder & 0.674 & 0.929 & 0.889 & 69.0 & 0.15 \tabularnewline 
    Siamese & 0.491 & 0.846 & 0.910 & 77.8 & 0.20 \tabularnewline 
    Triplets (exp) & 0.205 & 0.957 & 0.975 & 162.5 & \textbf{-0.03} \tabularnewline 
    Triplets (margin) & \textbf{0.194} & \textbf{0.967} & \textbf{0.977} & \textbf{33.7} & -0.04 \tabularnewline 
    \bottomrule
  \end{tabular}
\end{table}

The triplet-based channel charts obtain consistently better scores than the other considered approaches.
The key difference between the margin cost function in~\eqref{eq_triplets_margin} and the exponential cost function in~\eqref{eq_triplets_exp} lies in the $\mathsf{SR}$ metric.
The margin cost in this case leads to a chart that better respects the global shape of the original position.
Trustworthiness and continuity metrics show that they both maintain the neighborhoods, and the $\mathsf{EV}$ metric also indicates that the local density is also similar between the charts.
Interestingly, the representation obtained by the Siamese architecture appears very similar to that obtained through principal component decomposition of the input samples.

\subsubsection{Dataset 2}
In this dataset, a pedestrian user repeatedly walks an almost circular trajectory for 20 minutes (240000 samples), revolving over 3 times around the building at the center of the map (see Fig.~\ref{fig:circles_GNSS}).
The resulting charts are depicted in Figs.~\ref{fig:circles_PCA}--\ref{fig:circles_triplet_margin}, whereas the corresponding metrics are provided in Table~\ref{tab:circles}.
  
\begin{table}[hb!]
  \caption{Metric comparisons for different charting methods applied to Dataset 2.}
  \label{tab:circles}
  \centering
  \begin{tabular}{rrrrrr}
      \toprule
      Method &  \multicolumn{1}{c}{$\mathsf{KS}$} & \multicolumn{1}{c}{$\mathsf{TW}$} & \multicolumn{1}{c}{$\mathsf{CT}$} & \multicolumn{1}{c}{$\mathsf{SR}$} & \multicolumn{1}{c}{$\mathsf{EV}$} \tabularnewline
      \midrule
      MDS & 0.401 & 0.789 & 0.863 & 34.7 & 0.56 \tabularnewline 
      UMAP & 0.360 & 0.975 & 0.972 & 72.3 & 0.38 \tabularnewline 
      Autoencoder & 0.428 & 0.937 & 0.916 & 37.4 & 0.63 \tabularnewline 
      Siamese & 0.887 & 0.723 & 0.812 & $\gg$ 100 & $\gg$ 1 \tabularnewline 
      Triplets (exp) & \textbf{0.359} & 0.969 & 0.979 & \textbf{29.2} & \textbf{-0.03} \tabularnewline 
      Triplets (margin) & 0.404 & \textbf{0.985} & \textbf{0.985} & 39.9 & 0.04 \tabularnewline 
  \bottomrule
  \end{tabular}
\end{table}

Remarkably, Fig.~\ref{fig:circles_triplet_margin} shows that the triplet-based approach is the only one capable of correctly identifying the circular trajectories in the dataset---and in fact does so for both the exponential and margin cost function, although only the latter is shown on the figure because of space constraints.
This indicates that the metric learned by the network is globally relevant, although it is learned using only local information in the form of the proximity in time of the measured samples, as discussed in Section~\ref{sec:training_qualitative_information}.
The margin cost function~\eqref{eq_triplets_margin} even allows to distinguish on the chart two alternative and slightly different paths taken by the user, as depicted at the lower right of Fig.~\ref{fig:circles_GNSS}.
Also note that the triplet based network was trained with $T_f=\infty$.
This shows that the proposed algorithm is robust to samples in the training set being far in time but close in geographical position.
Table~\ref{tab:circles} shows a similar behavior as for dataset 1: triplet-based channel charting outperforms the alternative charting approaches according to all considered metrics.
The margin cost function once again performs well in both trustworthiness and continuity, losing slightly in all other metrics on this dataset.
We will use it as the reference cost function for triplet-based approaches in the following discussions and simulations.
The UMAP algorithm, while performing well both visually and in terms of numerical metrics, fails to properly identify the circular trajectories in this dataset.

\section{Extensions of the Triplet-based Approach}
\label{sec:extensions}

In this section, we discuss variations of the original architecture that can improve the performance and usefulness of triplet-based channel charting.

\begin{figure*}[ht]
  \centering
  \includegraphics[width=1\columnwidth]{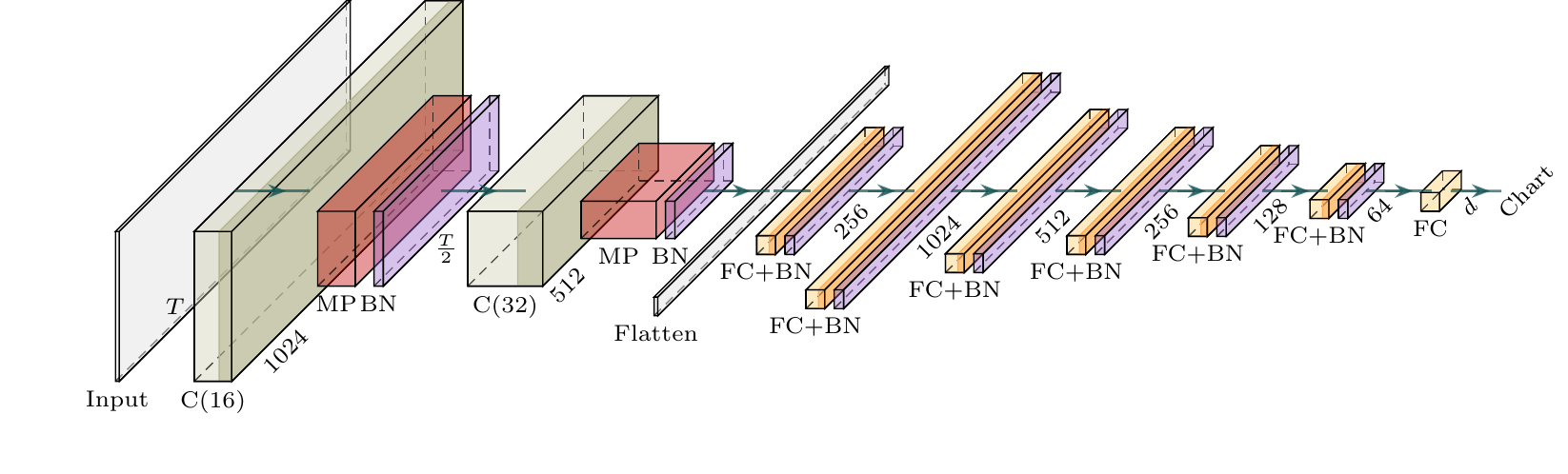}
  \vspace{-16pt}
  \caption{Neural network composed of convolutional layers with $F$ filters of size (3, 3) (C-$F$), max-pooling layers with (2, 2) filters (MP), fully connected layers (FC) and batch normalization (BN) layers. Each fully connected and convolutional layer except the last one apply a ReLU nonlinearity. The convolutional and max-pooling layers act on the time and frequency dimension.}
  \label{fig:conv_charting_nn}
\end{figure*}

\subsection{Curriculum Learning}
\label{sec:curriculum}

A key ingredient of the triplet approach actually relates to the construction of the training data set $\mathcal T$.
The number of possible choices of triplets in $S^3$ can however get very large, and $\mathcal T$ will likely be a small subset thereof.
Furthermore, the construction of $\mathcal T$ is intuitively an important parameter since there is not \emph{a priori} representative distribution of the triplets, and therefore the construction of the training batches will strongly influence the resulting channel chart.
This particular fact was identified in~\cite{Schroff2015,Hermans2017}, where various ways to increase the accuracy and training speed of the triplet-based approaches were discussed.
The selection of triplets can also be made increasingly challenging so that the embedding discriminates better between the classes; such an approach is called \emph{curriculum learning} or \emph{shaping}~\cite{Bengio2009}.

For the particular case at hand, the selection of the training set lies in choosing the values for $T_c$ in~\eqref{eq:triplet_time_threshold}, and possibly bounding the range of the \emph{far} sample with $T_f$ (see Fig.~\ref{fig:curriculum_thresholds}).
A natural curriculum in this setup would be to decrease $T_c$ and $T_f$ over the training epochs following a predetermined rate.
This reduction acts as an additional parameter which has to be optimized.
The results of this experiment are summarized in Table~\ref{tab:curriculum}.

\begin{table}[hb!]
  \caption{Metrics for the curriculum learning with differents values for $T_c$ and $T_f$, trained over 100 epochs on a 45 minutes (2700 seconds) subset of Dataset 1 with a margin loss.}
  \label{tab:curriculum}
  \centering
  \begin{tabular}{ccrrrrr}
    \toprule
    $T_c$ (s) & $T_f$ (s) & \multicolumn{1}{c}{KS} & \multicolumn{1}{c}{TW} & \multicolumn{1}{c}{CT} & \multicolumn{1}{c}{SR} & \multicolumn{1}{c}{EV} \tabularnewline
    \midrule
    0.2 & 2400 & \textbf{0.198} & \textbf{0.975} & \textbf{0.982} & \textbf{24.4} & 0.24 \tabularnewline 
    0.2 & 600 & 0.244 & 0.973 & 0.979 & 26.1 & \textbf{-0.11} \tabularnewline 
    0.2 & 2400 $\to$ 600 & 0.213 & 0.973 & 0.980 & 131.7 & 0.25 \tabularnewline 
    0.2 & 2400 $\to$ 60 & 0.220 & 0.967 & \textbf{0.982} & 141.1 & 0.20 \tabularnewline 
    \midrule
    0.5 $\to$ 0.05 & 2400 & 0.212 & 0.962 & 0.975 & 107.4 & 0.30 \tabularnewline 
    0.5 $\to$ 0.05 & 600 & 0.208 & 0.970 & 0.978 & 141.5 & 0.17 \tabularnewline 
    0.5 $\to$ 0.05 & 2400 $\to$ 600 & 0.218 & 0.963 & 0.976 & 63.8 & 0.14 \tabularnewline 
    0.5 $\to$ 0.05 & 2400 $\to$ 60 & 0.230 & 0.963 & 0.977 & 34.1 & 0.18 \tabularnewline 
  \bottomrule
  \end{tabular}
\end{table}

We consider here a 45 minute subset (2700 seconds) of Dataset 1, and set an upper bound of 2400 seconds on $T_f$.
It appears that the finer selection of the training set does not increase the quality of the obtained chart.
Using a smaller value for $T_f$ does not impact the metrics except the \textsf{EV}.
Similarly, reducing $T_f$ over the training epochs does not substantially improve the chart quality.
We therefore keep a fixed value for $T_c = 0.2$ and $T_f = 2400$ for the remainder of our analysis.

\subsection{Using Consecutive Time Samples to Build the Chart}
\label{sec:consecutive_samples}
So far, the time dimension was not exploited except for triplet selection.
This section considers aggregating consecutive samples in a sliding window fashion, in order to capture information related to the time evolution of the input samples.
Meta-samples are constructed by concatenating a number $L$ of consecutive samples:
\begin{equation}
  \bm {\tilde X}_i (\cdots) = \begin{pmatrix}
    \bm x_i (\cdots) \\
    \bm x_{i-1} (\cdots) \\
    \vdots\\
    \bm x_{i-L} (\cdots) \\
  \end{pmatrix}
  \label{eq:consecutive_samples_in_features}
\end{equation}
Note that meta-samples in~\eqref{eq:consecutive_samples_in_features} are potentially overlapping; in other words, we do not limit $i$ to be an integer multiple of $L$.
In this section the quality of the channel charts obtained using a dense neural network as in Fig.~\ref{fig:localization_nn} is compared to that obtained through a more complex network, depicted in Fig.~\ref{fig:conv_charting_nn}.
The latter contains 2 convolutional layers followed by max-pooling layers in the time and frequency dimension of the input samples in the first layers; the second part of the network is densely connected.
Each convolutional layer uses a $(3, 3)$ 2-dimensional kernel over the time and frequency axes.
The first convolutional layer generates 16 filters, and the second 32 filters.
Each max-pooling layer uses a $(2, 2)$ 2-dimensional kernel over the time and frequency axes.
Table~\ref{tab:consecutive} summarizes the performance of both networks when applied to meta-samples formed by a variable number of consecutive samples, with  $L$ ranging from 5 to 20.
The networks are trained on Dataset 1 using similar parameters and setups as described in Section~\ref{sec:experimental_results}.

\begin{table}[hb!]
  \caption{Metric comparisons for charting using the meta-samples \eqref{eq:consecutive_samples_in_features}. The ``dense'' network refers to the one in Section~\ref{sec:nn_building_block} and in Fig.~\ref{fig:localization_nn}, whereas ``convolutional'' refers to the network considered in this section and in Fig.~\ref{fig:conv_charting_nn}.}
  \label{tab:consecutive}
  \centering
  \begin{tabular}{ccccccc}
    \toprule
    Network & $L$ & \multicolumn{1}{c}{KS} & \multicolumn{1}{c}{TW} & \multicolumn{1}{c}{CT} & \multicolumn{1}{c}{SR} & \multicolumn{1}{c}{EV} \tabularnewline
    \midrule
    Dense & 5 & 0.223 & 0.959 & 0.973 & 126.7 & \textbf{-0.04} \tabularnewline 
    Dense & 10 & 0.214 & 0.963 & 0.976 & 144.4 & -0.05 \tabularnewline 
    Dense & 20 & 0.215 & 0.948 & 0.969 & 27.8 & -0.06 \tabularnewline 
    Convolutional & 5 & \textbf{0.158} & 0.969 & 0.979 & \textbf{20.5} & -0.05 \tabularnewline 
    Convolutional & 10 & 0.191 & \textbf{0.974} & 0.979 & 29.6 & -0.08 \tabularnewline 
    Convolutional & 20 & 0.231 & 0.972 & \textbf{0.980} & 29.3 & -0.05 \tabularnewline 
  \bottomrule
  \end{tabular}
\end{table}

\begin{figure}[ht!]
  \centering
  \includegraphics[width=0.4\columnwidth]{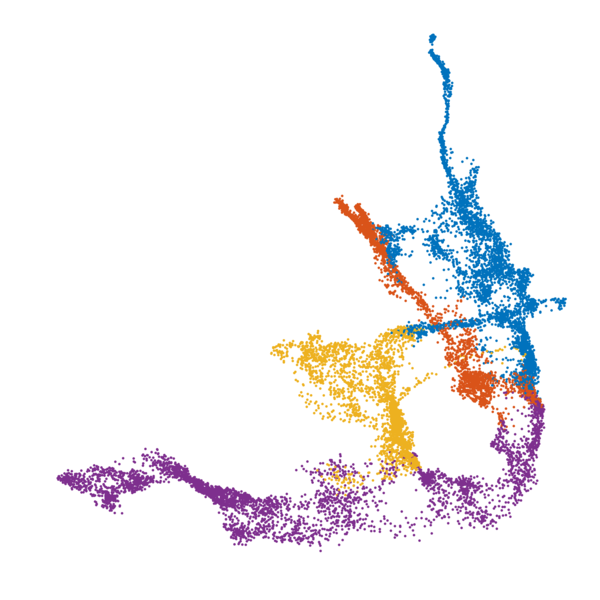}
  \caption{Chart obtained using triplet-based charting and meta-samples \eqref{eq:consecutive_samples_in_features} with $L=5$ consecutive samples. The inner neural network used for this chart is the convolutional network described in Fig.~\ref{fig:conv_charting_nn}.}
  \label{fig:consecutive}
\end{figure}

Using consecutive samples and the convolutional neural network of Fig.~\ref{fig:conv_charting_nn} provides a pronounced gain in both neighborhood-type metrics.
Fig.~\ref{fig:consecutive} also shows that the resulting chart does appear to capture better some elements of the true position, such as the thin corridors in the upper right part of Fig.~\ref{fig:pedestrian_GNSS}.
However, this comes at a complexity cost both in training and in the subsequent use of the trained network.
Indeed, the number of parameters increases to more than 5 millions in the network of Fig.~\ref{fig:conv_charting_nn}, despite having a bottleneck at the beginning of the fully connected layers.

\subsection{Semi-supervised Training}
\label{sec:semi_supervised_training}
As discussed in Section~\ref{sec:dimensionality_reduction}, one possible use of a channel chart is to act as a pseudo-map, onto which the user's positions can be tracked and estimated from its CSI.
In the self-supervised case considered so far, this pseudo-map lives in an arbitrary latent space.
However, if some information about the true geographical position of the user is available (through some labeled samples) during the chart construction, semi-supervised methods allow to link the chart to an actual geographical map.
Let us assume that position labels $\bm p_i \in \mathbb R^2$---or $\bm p_i \in \mathbb R^3$---are known for some samples; we let $\mathcal P \subset \mathcal S$ denote the set of indices for which position labels are available.
If the latent space dimension is the same dimension as the position, the cost function \eqref{eq_triplets} can be adapted to the semi-supervised scenario by adding a penalty term proportional to $\|\bm p_i - \bm y_i\|$ for the samples in $\mathcal P$.
For arbitrary latent space dimensions, this can be generalized to a constraint that a projection of the latent vector $\bm y_i$ is equal to the position $\bm p_i$.
The margin cost in~\eqref{eq_triplets_margin} is thus replaced by
\begin{equation}
\frac{1}{|\mathcal T|} \sum_{(i, j, k)\in\mathcal T}  \Big(d_{\bm \theta}(\bm x_i, \bm x_j) - d_{\bm \theta}(\bm x_i, \bm x_k) + M\Big)^+ 
 + \frac{\alpha}{|\mathcal P|} \sum_{i \in \mathcal P}  \left\| \bm P f_{\bm \theta}( \bm x_i ) - \bm p_i \right\|\label{eq_triplets_semisupervised}
\end{equation}
where $\bm P$ is a matrix in $\mathbb R^{2 \times d}$, and $\alpha$ is a parameter weighting the penalty associated with the position-labeled samples.
The matrix $\bm P$ can be either learned along with the rest of the neural network, or chosen to be fixed to e.g. force only a subset of the coordinates in the output vector to match the supervised position information.

In all experiments on semi-supervised learning, the networks were trained over 100 epochs on Dataset 1 with a margin loss, using the Adam optimizer.
The dimension of the latent space is here again set to $d=2$, and the matrix $\bm P$ in \eqref{eq_triplets_semisupervised} is fixed and chosen as the identity matrix of size 2.
The labeled samples are assumed to be regularly spread over time in the training set; for instance, 9 known positions over 45 minutes would be assumed to occur at a 5 minutes interval.
Points within $\pm$ 0.5 seconds of the known positions are assumed to have the same position.
The close samples bound $T_c$ is set to 0.2 seconds, and the far sample bound $T_f$ to 2400 seconds.
For each mini-batch used for training, 1\% of the anchors are selected to be in the known-position set.
We also compare the relative performance of this semi-supervised extension to representation-constrained auto-encoders~\cite{huang2019charting_representation_constrained}.
We apply the fixed absolute distance (FAD) regularizer from~\cite{huang2019charting_representation_constrained} with anchors chosen as explained above.
As with the triplet cost~\eqref{eq_triplets_semisupervised}, the FAD regularizer for the representation-constrained autoencoders is weighted with a parameter $\alpha$.

\begin{table}[hb!]
\caption{Metrics for the semi-supervised extension of the charting approach.}
\label{tab:semi-supervised}
\centering
\begin{tabular}{cccrrrrr}
  \toprule
  $\alpha$ & $|\mathcal P|$ & $M$ & \multicolumn{1}{c}{KS} & \multicolumn{1}{c}{TW} & \multicolumn{1}{c}{CT} & \multicolumn{1}{c}{SR} & \multicolumn{1}{c}{EV} \tabularnewline
  \midrule
  \multicolumn{8}{c}{Triplet-based network} \tabularnewline
  \midrule
  1 & 9 & 0.2 & 0.299 & 0.944 & 0.968 & 40.5 & 0.40 \tabularnewline 
  1 & 45 & 0.2 & 0.195 & \textbf{0.952} & 0.969 & 22.8 & 0.53 \tabularnewline 
  10 & 9 & 0.2 & 0.364 & 0.918 & 0.945 & 45.8 & 0.14 \tabularnewline 
  10 & 45 & 0.2 & 0.279 & 0.931 & 0.944 & 30.4 & 0.71 \tabularnewline 
  1 & 9 & 1 & 0.256 & 0.946 & 0.970 & 34.4 & 0.25 \tabularnewline 
  1 & 45 & 1 & \textbf{0.184} & \textbf{0.952} & \textbf{0.971} & \textbf{22.5} & 0.63 \tabularnewline 
  10 & 9 & 1 & 0.385 & 0.919 & 0.943 & 49.4 & 0.15 \tabularnewline 
  10 & 45 & 1 & 0.255 & 0.929 & 0.948 & 29.2 & 0.52 \tabularnewline 
  \midrule
  \multicolumn{8}{c}{Representation-constrained auto-encoders~[8]} \tabularnewline
  \midrule
  1 & 9 & - & 0.573 & 0.927 & 0.892 & 84.6 & 0.15 \tabularnewline 
  1 & 45 & - & 0.249 & 0.937 & 0.935 & 30.3 & \textbf{0.09} \tabularnewline 
  10 & 9 & - & 0.670 & 0.877 & 0.879 & 84.2 & 0.10 \tabularnewline 
  10 & 45 & - & 0.270 & 0.937 & 0.935 & 29.3 & 0.10 \tabularnewline 
  \bottomrule
\end{tabular}
\end{table}

Different values for the meta-parameters in our models and cost function are considered.
In particular, the margin $M$ in~\eqref{eq_triplets_margin} may become a relevant parameter in the semi-supervised case, whereas for the self-supervised case it was arbitrary due to the invariance of the optimization function with respect to global scaling.
The cost function indicates that the margin is homogeneous to a distance in the latent space.
The value for the margin may thus be linked \emph{a priori} to the global scale of the constraint and the value chosen for $T_c$, taking into account the velocity of the user during the measurement.
The data was gathered at pedestrian speed in this dataset, therefore the average speed can be assumed close to 1 m.s\textsuperscript{-1}.
We therefore compare having the margin kept at the value 1, and setting it as $M = T_c$ and assuming a speed of 1 m.s\textsuperscript{-1}.
We also consider a stronger penalty for the known position in the form of a larger value for $\alpha$, and having known points every 5 minutes or every minute in the training dataset.
\begin{figure}[ht!]
  \centering
  \includegraphics[width=0.4\columnwidth]{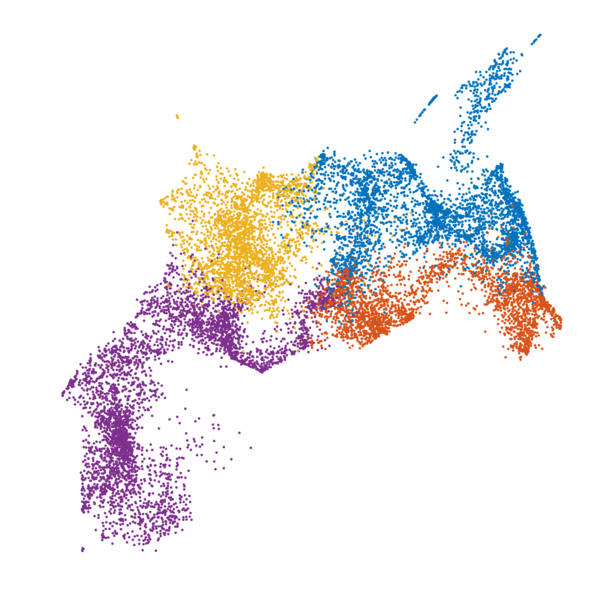}
  \caption{Chart obtained using triplet-based charting for the semi-supervised case, using the cost function in~\eqref{eq_triplets_semisupervised} with $\alpha = 1$ and $M = 1$. There are $|\mathcal P| = 45$ known points in the training dataset of 45 minutes.}
  \label{fig:semisupervised}
\end{figure}
The numerical metrics are reported in Table~\ref{tab:semi-supervised}.
The neighborhood scores---the TW and CT metrics---decrease slightly in the semi-supervised results.
As depicted in Fig.~\ref{fig:semisupervised}---to be compared to the real position of Fig.~\ref{fig:pedestrian_GNSS}---the chart also appears to have a higher local variance.
Note that the decrease in performance according to the considered metrics (which measure the quality of the dimensionality reduction) is expected, since adding the constraint on the anchors essentially turns the dimensionality reduction problem into a more difficult multi-objective problem.
From the excess variance scores, it seems that the representation-constrained auto-encoders perform better on this front.
Table~\ref{tab:semi-supervised} also indicates that changing the value of the margin $M$ does not appear to have a significant effect on the overall performance.
Furthermore, when less anchor points are available with respect to the size of the dataset, increasing the weight related to the anchor points in the cost function reduces the excess variance.

\section{Conclusion}

In this paper, we introduced a novel approach to channel charting based on triplets of samples.
The algorithm uses the timestamp information naturally available in wireless channel CSI samples to create a channel chart.
We showed that the performance of the proposed technique is competitive with respect to the state-of-the-art in channel charting, both visually and quantitatively with respect to classical metrics.
We then studied different extensions of the approach, in particular for semi-supervised learning.

Future research in this direction could include different ways to make use of the timestamp information.
Siamese networks and auto-encoders could in particular be readily augmented to consider it.
This might offer alternatives to the triplet-based approach.
Another potentially interesting avenue of research lies in online learning of channel charts.
In particular, in order to make channel charting usable in practice, one would require base stations to build their channel representation on the fly, and change them over time to adapt to modification of the environment.
Although we have shown in~\cite{Localization_Globecom} that the networks trained in a realistic deployment setup enjoy some stability over time, it still requires an adequate training dataset to achieve such results.

\bibliographystyle{IEEEtran}
\bibliography{charting-jsac}

\end{document}